\documentclass[showpacs,amsmath,amssymb,aps,10pt,reprint,superscriptaddress,longbibliography]{revtex4-1}
\usepackage{graphicx}
\usepackage{bm}
\usepackage[breaklinks=true,colorlinks=true,linkcolor=blue,urlcolor=blue,citecolor=blue]{hyperref}
\usepackage{graphics}
\usepackage{dcolumn}
\usepackage{amsmath,amssymb}
\usepackage{mathdots}
\usepackage{natbib}
\usepackage{soul,color}
\usepackage{epstopdf}
\usepackage[note-name]{notes2bib}

\begin{document}
\title{\Large Supersonic dynamics of guided magnetic flux quanta}

\author{O.~V.~Dobrovolskiy}
    \email[Corresponding author: ]{Dobrovolskiy@Physik.uni-frankfurt.de}
    \affiliation{Physikalisches Institut, Goethe University, 60438 Frankfurt am Main, Germany}
    \affiliation{Physics Department, V. Karazin Kharkiv National University, 61022 Kharkiv, Ukraine}
\author{V.~M.~Bevz}
    \affiliation{Physics Department, V. Karazin Kharkiv National University, 61022 Kharkiv, Ukraine}
\author{E. Begun}
    \affiliation{Physikalisches Institut, Goethe University, 60438 Frankfurt am Main, Germany}
\author{R. Sachser}
    \affiliation{Physikalisches Institut, Goethe University, 60438 Frankfurt am Main, Germany}
\author{R. V. Vovk}
    \affiliation{Physics Department, V. Karazin Kharkiv National University, 61022 Kharkiv, Ukraine}
\author{M.~Huth}
    \affiliation{Physikalisches Institut, Goethe University, 60438 Frankfurt am Main, Germany}
\date{\today}

\begin{abstract}
The dynamics of Abrikosov vortices in superconductors is usually limited to vortex velocities $v\simeq1$\,km/s above which samples abruptly transit into the normal state. In the Larkin-Ovchinnikov framework, near the critical temperature this is because of a flux-flow instability triggered by the reduction of the viscous drag coefficient due to the quasiparticles leaving the vortex cores. While the existing instability theories rely upon a uniform spatial distribution of vortex velocities, the measured (mean) value of $v$ is always smaller than the maximal possible one, since the distribution of $v$ never reaches the $\delta$-functional shape. Here, by guiding magnetic flux quanta at a tilt angle of $15^\circ$ with respect to a Co nanostripe array, we speed up vortices to supersonic velocities. These exceed $v$ in the reference as-grown Nb films by almost an order of magnitude and are only a factor of two smaller than the maximal vortex velocities observed in superconductors so far. We argue that such high $v$ values appear in consequence of a collective dynamic ordering when all vortices move in the channels with the same pinning strength and exhibit a very narrow distribution of $v$. Our findings render the well-known vortex guiding effect to open prospects for investigations of ultrafast vortex dynamics.
\end{abstract}
\maketitle

\section{Introduction}
The dynamics of Abrikosov vortices in superconductors is usually restricted to vortex velocities $v \backsimeq 1$\,km \cite{Lar75etp,Lar86inb,Bez92pcs,Shk17prb,Mus80etp,Vol92fnt,Xia99prb,Per05prb,Sil12njp,Gri15prb} due to the presence of normal excitations (quasiparticles) in the vortex cores. When the electric field induced by vortex motion raises the quasiparticle energy above the potential barrier associated with the order parameter around the vortex core, quasiparticles may leave it and the vortex core shrinks. The shrinkage of the vortex core leads to a drastic reduction of the viscous drag coefficient and a further avalanche-like acceleration of vortices, eventually quenching the low-resistive state. In this Larkin-Ovchinnikov (LO) instability framework \cite{Lar75etp,Lar86inb,Bez92pcs}, the vortex instability velocity $v^\ast$ is related to the quasiparticle relaxation time
\begin{equation}
\label{e1}
    \tau_\mathrm{qp} = D[14\zeta(3)]^{1/2}[1-T/T_c]^{1/2}/(\pi v^{\ast2}),
\end{equation}
where $D$ is the quasiparticle diffusion coefficient and $\zeta(x)$ is the Riemann function. Thus, the relaxation of quasiparticles is pivotal in almost all phenomena harbouring non-equilibrium superconductivity \cite{Gra81boo,Siv03prl,Leo11prb,Cir11prb,Wor12prb,Bec13prl,Vis14prl,Che14nph,Lar15nsr,Dob19rrl}. In particular, it governs triggering vortex avalanches and dc-assisted microwave quenching in transmission lines \cite{Che14apl,Lar17pra,Loe19acs}, and is crucial for photon detection \cite{Had09nph,Cap17apl,Vod17pra} and optical control of dynamical states \cite{Mad18sca}. Therefore, approaches for probing the ultimate limits of $v^\ast$ are of fundamental importance as well as they are of interest for various applications.

Recently, pinning effects on the flux-flow instability have attracted attention experimentally \cite{Gri12apl,Sil12njp,Dob17sst} and theoretically \cite{Sil12njp,Gri15prb}. In particular, these studies have revealed a decrease of $v^\ast$ with increase of the pinning strength. In the presence of strong uncorrelated pinning the distribution of $v$ in the vortex ensemble has been argued to be broader than in the case of weak pinning \cite{Sil12njp}. Specifically, in samples with a narrower distribution of $v$ the measured (mean) value of $v_m^\ast$ is expected to be closer to the maximum possible $v^\ast$ value. In the case of periodic pinning, there has been a hint that $v^\ast$ attains a maximum when the spacing between vortex rows is commensurate with the period of the pinning potential \cite{Dob17sst}. Accordingly, efficient approaches to probe ultimate vortex velocities (and relaxation times) should include tailoring of the distribution of $v$ into a close-to-$\delta$-functional shape. In addition, faster relaxation processes are expected when a superconductor is put in contact with a ferromagnetic layer\cite{Tan07prb,Att12pcm,Cap17apl}.

In a recent study \cite{Bez19arx} of the effect of random disorder on $v^\ast$, a nearly twofold increase of $v^\ast$ has been reported for Nb films with flat morphology and weak pinning as compared to Nb films with grained morphology where pinning was strong. The experimental data were in agreement with the recently generalized LO theory for the case of local flux-flow instability \cite{Bez19arx} for which instability jumps occur in linear sections of the $I$-$V$ curves.

Here, we study the effect of correlated disorder on the flux-flow instability in superconducting Nb / ferromagnetic Co hybrid structures in which the Co layer is either continuous or periodically interrupted. In the latter case, by guiding magnetic flux quanta at a tilt angle of $15^\circ$ with respect to a Co nanostripe array, we speed up vortices to supersonic velocities. These exceed $v^\ast$ in as-grown Nb films and Nb/Co bilayers by an order of magnitude and are only a factor of two smaller than the maximal vortex velocities ever observed \cite{Emb17nac}.

\section{Experimental}
\begin{figure}[t!]
    \centering
    \includegraphics[width=1\linewidth]{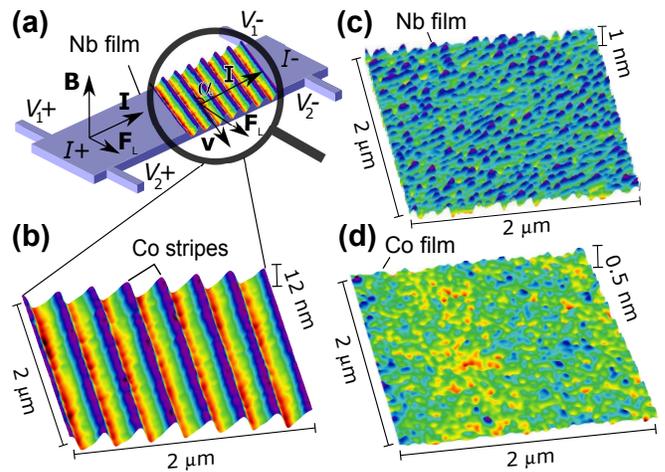}
    \caption{(a) Experimental geometry. The transport current $\mathbf{I}$ in a superconducting film exerts the Lorentz force $\mathbf{F}_\mathrm{L} = \mathbf{I}\times\mathbf{B}$ on Abrikosov vortices.
    The Co stripes deposited on top of the Nb film under a tilt angle $\alpha$ with respect to $\mathbf{I}$ induce pinning channels for vortices and make them to move with the velocity $\mathbf{v}\nparallel \mathbf{F}_\mathrm{L}$ due to the vortex guiding effect. Longitudinal $V^+_\parallel=V_2^- - V_2^+$ and transverse $V^+_\perp=V_2^+ - V_1^+$ voltages are measured as a function of the current value.
    Atomic force microscopy images are shown for a Nb film decorated with Co stripes (b), an as-grown Nb film (c), and a continuous Co layer (d) deposited on top of a Nb film. }
    \label{f1}
\end{figure}

The non-equilibrium state generated by high-velocity vortex motion was studied in Nb/Co hybrid structures and one as-grown Nb film. The epitaxial (110) Nb films with a thickness of $70$\,nm were prepared by dc magnetron sputtering on a-cut sapphire substrates. During the deposition, the argon pressure was $4\times10^{-3}$\,mbar, the substrate temperature was $850^\circ$C, and the deposition rate was $1$\,nm/s \cite{Dob12tsf}. The films were pre-patterned by photolithography and Ar ion-beam etching in order to define $10\,\mu$m$\times 100\,\mu$m structures for electrical resistance measurements. The experimental geometry is shown in Fig. \ref{f1}(a).

In the hybrid structures, either a $10$\,nm-thick continuous Co layer or an array of Co stripes was deposited on top of the Nb film by focused electron beam induced deposition (FEBID). FEBID represents a direct-write process by which a metal-organic precursor gas, in this case Co$_2$(CO)$_8$, adsorbed on a film surface, is dissociated in the focus of the electron beam into a permanent deposit and volatile components \cite{Hut18mee}. FEBID was done in a high-resolution scanning electron microscope FEI Nova NanoLab 600 with the beam parameters $10$\,kV/$2.1$\,nA. The Co nanostripe array has a period $a$ of $300$\,nm, the stripes have a height of $12$\,nm and a half-height width of about $60$\,nm. The typical material composition in the Co-FEBID is $72$\,at\% of Co, $12$\,at\% of C, and $16$\,at\% of O, as deduced from energy-dispersive x-ray spectroscopy on thicker layers deposited with the same beam parameters \cite{Beg15nan}. Non-contact atomic force microscopy images of the samples' surfaces are shown in Fig. \ref{f1}(b)-(d).

The as-grown Nb films are superconducting below $T^\mathrm{Nb}_\mathrm{c} = 8.98$\,K, as defined by the $50$\% resistance drop criterion, while $T^\mathrm{Nb/Co}_\mathrm{c}=8.41$\,K and $T^\mathrm{Nb/Co\,stripes}_\mathrm{c} \approx 8.67$\,K. In the hybrid structures, the suppression of $T_\mathrm{c}$ is attributed to the proximity effect \cite{Buz05rmp,Kom14apl}.
The upper critical field $H_\mathrm{c2}(0)$ of the as-grown Nb film is about $1.12$\,T, as deduced from fitting the dependence $H_\mathrm{c2}(T)$ to the standard expression $H_\mathrm{c2}(T) = H_{c2}(0) [1-(T/T_\mathrm{c})^2]$. In the Nb/Co bilayer $H_\mathrm{c2}(0) \simeq 1.28$\,T and $H_\mathrm{c2}(0) \simeq 1.42$\,T in the decorated Nb film. From the relation $H_\mathrm{c2}(0) = [\Phi_0/(2\pi H_\mathrm{c2}(0))]^{1/2}$ the superconducting coherence length $\xi(0)$ was estimated to be around $15$\,nm for all samples. Near $T_c$, temperature dependences of the upper critical field can be fitted to straight lines with the slopes $dB^\mathrm{Nb}_{\mathrm{c}2}/dT = 0.24$\,T/K, $dB^\mathrm{Nb/Co}_{\mathrm{c}2}/dT = 0.33$\,T/K, and $dB^\mathrm{Nb/Co\,stripes}_{\mathrm{c}2}/dT = 0.28$\,T/K. These correspond to the electron diffusivity coefficients $D^\mathrm{Nb} = 4.57$\,cm$^2$/s, $D^\mathrm{Nb/Co} = 3.32$\,cm$^2$/s, and $D^\mathrm{Nb/Co\,stripes} = 3.97$\,cm$^2$/s, respectively, as deduced from the relation $D = -1.097(dB^\mathrm{Nb}_{\mathrm{c}2}/dT)^{-1}|_{T = T_\mathrm{c}}$ \cite{Gui86ltp}.

The $I$-$V$ curves were recorded in the current-driven regime with magnetic fields oriented perpendicular to the sample surface. In the decorated Nb films, magnetic flux quanta more easily move along the Co stripes than overcome the barriers of the washboard pinning potential induced by the nano-array \cite{Dob17sst,Dob17pcs}. For this reason we present the measured voltages without Hall-like contributions \cite{Zec18prb,Sil10inb} arising in our samples due to the pinning anisotropy \cite{Shk06prb,Dob16sst} and discriminate the longitudinal $V^+_\parallel$ and transverse $V^+_\perp$ voltage components which are even with respect to magnetic field reversal, namely $V^+_{\parallel,\perp}(B) \equiv [V_{\parallel,\perp}(B) + V_{\parallel,\perp}(-B)]/2$.

\section{Results and discussion}
\begin{figure}[t!]
    \centering
    \includegraphics[width=1\linewidth]{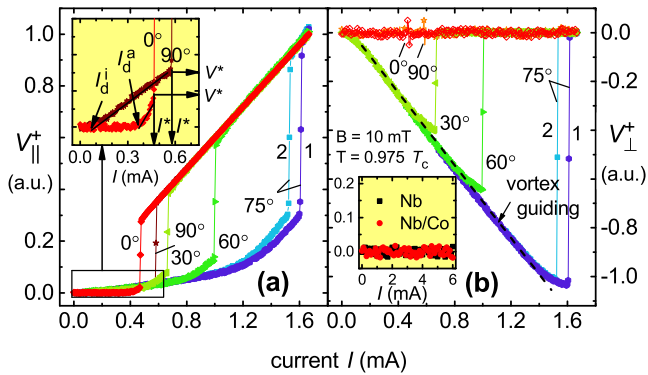}
    \caption{The $I$-$V$ curves for the longitudinal (a) and perpendicular (b) voltage components.
    Inset in (a) depicts the definition of the depinning currents $I^i_d$ and $I^a_d$, the instability current $I^\ast$, and the instability voltage $V^\ast$.
    Inset in (b) demonstrates the absence of $V^+_\perp$ in the as-grown Nb film and in the Nb/Co bilayer.
    The curves $1$ and $2$ were measured for the positive and negative current polarity, respectively.}
    \label{f2}
\end{figure}

The effect of the Co stripes tilted at different angles $\alpha$ with respect to the current direction on the vortex dynamics is illustrated in Fig. \ref{f2} where $I$-$V$ curves are shown for the $V^+_\parallel$ and $V^+_\perp$ components at $T = 0.975T_c$ and $B = 10$\,mT. Since the Co stripes increase the effective films' cross-section and thereby reduce their normal-state resistances by up to $5\%$, in Fig. \ref{f2}(a) $V^+_\parallel$ is normalized to the voltage at $1.6$\,mA while in Fig. \ref{f2}(b) $V^+_\perp$ is presented after normalization of the differential resistance to its value in the linear regime of vortex guiding, as depicted by the dashed line in Fig. \ref{f2}(b). Remarkably, with increase of $\alpha$ from $0^\circ$ to $75^\circ$ in Fig. \ref{f2}(a), a significant extension of the low-resistive state towards larger currents is observed. This is accompanied by the development of upturn bendings at the foots of the instability jumps. Furthermore, at lower currents, referring to the inset of Fig. \ref{f2}(a), the peculiar behavior of the $I$-$V$ curves at $\alpha = 0^\circ$ and $90^\circ$ should be noted. Specifically, if we introduce the depinning currents $I_\mathrm{d}^\mathrm{i}$ and $I_\mathrm{d}^\mathrm{a}$ for the intrinsic and Co-array-induced anisotropic pinning, respectively, we reveal that $I_\mathrm{d}^\mathrm{i} < I_\mathrm{d}^\mathrm{a}$ contrasts with the inequality $I^\ast(0^\circ) < I^\ast(90^\circ)$ for the instability currents. The definition of $I_\mathrm{d}^\mathrm{i}$, $I_\mathrm{d}^\mathrm{a}$, and $I^\ast$ is shown in the inset of Fig. \ref{f2}(a). Thus, the effect of the Co nanostripe array on the depinning and instability currents is opposite. Namely, whereas the strong correlated pinning results in an \emph{increase} of $I_\mathrm{d}$, it simultaneously leads to a \emph{reduction} of $I^\ast$. The reduction of $I^\ast$ is attributed to a broader distribution of vortex velocities as they are getting depinned from strong pinning sites \cite{Sil12njp}.

For the examination whether the instability jumps occur in the guiding regime with $\mathbf{v}\nparallel\mathbf{F}_\mathrm{L}$ or in the regime of free flux flow we turn to Fig. \ref{f2}(b) for $V^+_\perp$ whose nonzero value is a fingerprint of the vortex guiding effect \cite{Nie69jap,Shk06prb,Sil10inb,Dob16sst}. Specifically, from Fig. \ref{f2}(b) it follows that the instability jumps, which are seen at the same current values in both voltage components, occur in the vortex guiding regime at all $\alpha\neq 0^\circ,\,90^\circ$. Indeed, it is known that when vortices move parallel or perpendicular to pinning channels, no tangential component of the driving force arises which is responsible for the guiding effect \cite{Nie69jap,Shk06prb,Sil10inb,Dob16sst}. For completeness, the absence of $V^+_\perp$ in the as-grown Nb film and the Nb/Co bilayer is shown in the inset in Fig. \ref{f2}(b), thereby further corroborating that, in both reference samples, the instability occurs in the regime of free flux flow. We therefore conclude that the extension of the low-dissipative regime in the $I$-$V$ curves in Fig. \ref{f2}(a) at $\alpha\neq 0^\circ,\,90^\circ$ is caused by guided vortex motion and the stabilizing effect of vortex guiding on viscous flux flow is most pronounced at $\alpha = 75^\circ$, on which we will focus in the remainder of our presentation.

\begin{figure}[t!]
    \centering
    \includegraphics[width=1\linewidth]{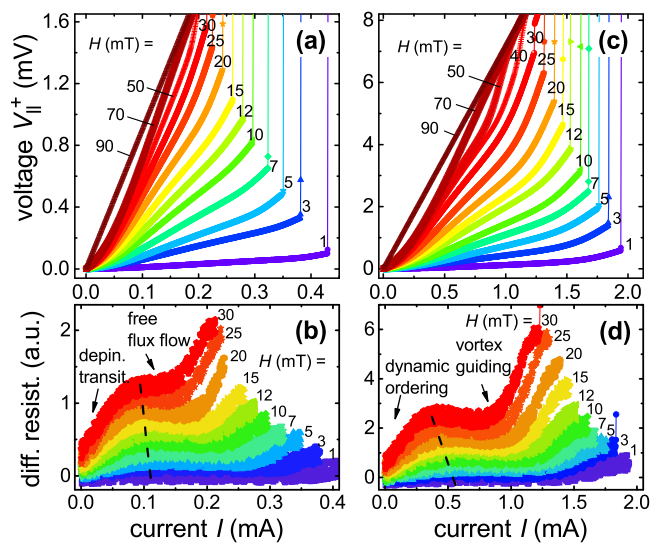}
    \caption{$I$-$V$ curves for the as-grown Nb film (a) and the Nb film decorated with Co stripes (c) with $\alpha = 75^\circ$ at $T = 0.975T_\mathrm{c}$ for a series of magnetic field values, as indicated.
    The respective differential resistances are shown in panels (b) and (d).}
    \label{f3}
\end{figure}

Figure \ref{f3} displays the $I$-$V$ curves for the Nb film with Co stripes and the as-grown Nb film at $T = 0.975T_c$. At larger currents, the $I$-$V$ curves exhibit abrupt transitions to the normal state in the magnetic field range $1$ to $30$\,mT. We note ahead that for the most part of these field values the vortex lattice is sparse with respect to the nanostripe array, that is the vortex lattice parameter $a_\triangle = (2\Phi_0/\sqrt{3}B)^{1/2}$ is larger than its period $a=300$\,nm. Here $\Phi_0$ is the magnetic flux quantum. While in principle one could expect commensurability effects at $B = 26.5$\,mT, corresponding to $a_\triangle =a$, previous studies revealed rather weak peculiarities at matching fields in Nb films decorated with Co-FEBID stripes \cite{Dob11pcs}. Furthermore, these peculiarities vanished with increase of the guiding angle.

The $I$-$V$ curves for the Nb/Co bilayer (not shown) and the Nb film, shown in Fig. \ref{f3}(a), are qualitatively similar. At small currents $I\lesssim 0.1$\,mA, there is a nonlinear transition to a regime with nearly constant differential resistance followed by an upward bending preceding the instability jumps. By contrast, the shape of the $I$-$V$ curves for the decorated Nb film in Fig. \ref{f3}(c) is noticeably different. The instability jumps occur at the ends of much more extended linear sections, and shallow minima can be recognized at $I\backsimeq1$\,mA in the $dV(I)/dI$ curves. We note that these smooth peaks followed by the dips in Fig. \ref{f3}(d) are reminiscent of the initial part of the $dV(I)/dI$ curves for NbN films with strong pinning \cite{Gri12apl}. The nonlinear transition at $I\lesssim 0.1$\,mA for the Nb film and the Nb/Co bilayer is attributed to variations of the individual Lorentz forces as vortices are getting depinned before the establishment of free flux flow \cite{Kos94prl}. By contrast, in the decorated Nb film, the nonlinear transition at $I\lesssim 0.5$\,mA is assumed to be caused by a dynamic  ordering of vortices \cite{Gri12apl} as they are guided by the Co stripes.
\clearpage

The decisive role of vortex guiding on the flux-flow instability can be corroborated even further if one decreases the strength of the anisotropic pinning while keeping the guiding angle constant. In our hybrid structure this can readily be achieved by reverting the current polarity (and the direction of the vortex motion). Namely, due to the edge steepness of the Co stripes pre-defined differently in the FEBID process, we refer to Fig. \ref{f1}(b), the $I$-$V$ data acquired under current polarity reversal exhibit slightly different $I^\ast$ values. This is because of the asymmetry of the pinning potential landscape induced by the stripes, which acts as a washboard ratchet potential \cite{Plo09tas,Sil10inb,Dob17nsr}. In this, when driven by a transport current of positive polarity, the driving force acting on vortices makes them to overcome the steep slopes of the Co stripes, that preserves a stable flux flow up to larger current values. Importantly, the $V^\ast$ value does not vary under current polarity reversal. This can be understood as a consequence of the independence of the quasiparticle relaxation time $\tau_\mathrm{qp}$ of the direction of the guided vortex motion.

\begin{figure}[t!]
    \centering
    \includegraphics[width=1\linewidth]{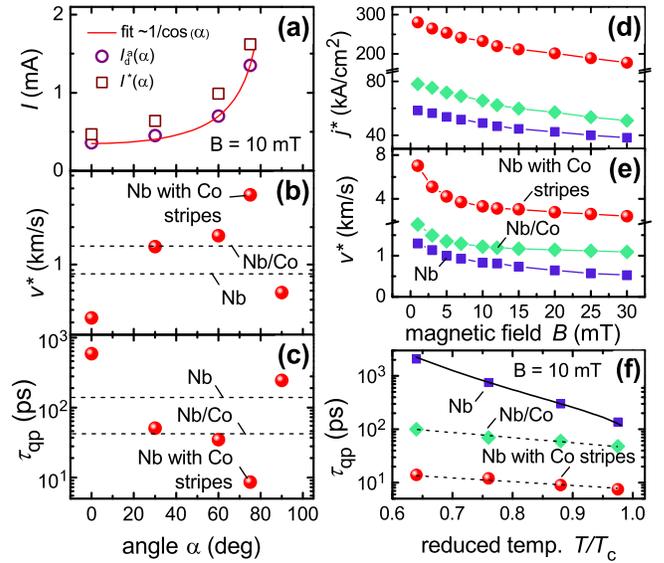}
    \caption{Instability $I^\ast$ and depinning $I_\mathrm{d}^\mathrm{a}$ currents, instability velocity $v^\ast$, and
    the relaxation time $\tau_{\mathrm{qp}}$ (c) deduced by Eq. \ref{e1} as a function of the guiding angle $\alpha$ at $B = 10$\,mT.
    The dashed lines in (b) and (c) depict the respective values for the Nb/Co bilayer and the as-grown Nb film.
    Dependences of the instability current density $j^\ast$ (d) and the instability velocity $v^\ast$ (e) on the magnetic field value.
    In panels (a)-(e) $T = 0.975 T_\mathrm{c}$.
    (f) Temperature dependence of the relaxation time at $B = 10$\,mT.
    The solid line is a fit $\tau^\mathrm{Nb}_\mathrm{qp} \propto \exp[2\Delta(T)/k_\mathrm{B}T]$ with $\Delta(0)\approx 1.85 k_\mathrm{B}T_\mathrm{c}$. The dashed lines are guides for the eye.}
    \label{f4}
\end{figure}

Proceeding to a quantitative analysis of the instability parameters we first note that at all angles $\alpha \leq 75^\circ$ the depinning current value nicely follows the expected dependence $I^\mathrm{a}_\mathrm{d}(\alpha)\sim I^\mathrm{a}_\mathrm{d}(0^\circ)/\cos\alpha$ as only the normal component of $\mathbf{F}_\mathrm{L}$ is responsible for overcoming the linearly-extended pinning barriers by vortices \cite{Shk06prb}. Here, $I^\mathrm{a}_\mathrm{d}(0^\circ)\simeq 0.4$\,mA corresponds to a depinning current density $j^\mathrm{a}_\mathrm{d}(0^\circ)$ of $60$\,kA/cm$^2$. The angle dependence of the instability current $I^\ast(\alpha)$ is illustrated in Fig. \ref{f4}(a) and appears to correlate with $I^\mathrm{a}_\mathrm{d}(\alpha)$. The correlation of both currents is indicative of the important role of the anisotropic pinning in the maintaining of the low-resistive state.

The angle dependence $v^\ast(\alpha)$ is non-monotonic. This is because of the different pinning and guiding regimes encountered upon the rotation of $\alpha$. Specifically, with increase of $\alpha$ from $0^\circ$ to $75^\circ$, $v^\ast$ first grows from about $300$\,m/s to above $3.3$\,km/s and then drops to $620$\,m/s at $90^\circ$. Here, $v^\ast(\alpha)$ has been calculated by the relation $v^\ast(\alpha) = V^\ast_\parallel/(B L\sin\alpha)$, where $L=100\,\mu$m is the distance between the voltage contacts and $\alpha\neq0^\circ$. This has been done to account for crossing of vortices of the superconducting constriction at an oblique angle. Obviously, while at $\alpha=75^{\circ}$ the geometric factor $\sin^{-1}(\alpha)$ results in a small increase of the full vortex velocity by about $3.5\%$ as compared to $v^\ast_\perp$ deduced from $V^+_\parallel$, the deduced $v^\ast(30^{\circ})$ and $v^\ast(60^{\circ})$ values are affected more strongly. Importantly, at $10$\,mT, the high value of $v^\ast(75^\circ)\simeq3.3$\,km/s caused by the guiding effect is significantly larger than $v^\ast_\mathrm{Nb}\simeq 830$\,m/s and $v^\ast_\mathrm{Nb/Co}\simeq 1.2$\,km/s which are in line with the typical $v^\ast$ values in high-quality Nb films and Nb/ferromagnet hybrid structures \cite{Per05prb,Leo11prb,Att12pcm}. The dependences of $v^\ast(B)$ and $j^\ast(B)$ are presented in Fig. \ref{f4}(d) and (e). One sees that $v^\ast(75^\circ) \gtrsim 4$\,km/s are observed at $B \lesssim 5$\,mT, while $v^\ast(75^\circ) \gtrsim 2.4$\,km/s at all studied magnetic field values. The instability current density $j^\ast$ decreases with increase of $B$, as predicted \cite{Lar86inb,Bez92pcs} and most commonly observed \cite{Mus80etp,Vol92fnt,Sil12njp,Gri12apl,Dob17sst}, although other dependences are also known in the literature \cite{Per05prb,Sil12njp}.

Figure \ref{f4}(c) depicts the evolution of the quasiparticle relaxation time $\tau_\mathrm{qp}(\alpha)$ given by Eq. (\ref{e1}) at $B = 10$\,mT and $T = 0.975T_\mathrm{c}$. Since $\tau_\mathrm{qp} \propto D/(v^\ast)^2$ the quasiparticle relaxation time for the guided motion of vortices along the Co stripes reaches $\tau_\mathrm{qp}^\mathrm{Co\,stripes}(75^\circ) \simeq 10$\,ps. This is by about one order of magnitude smaller than  $\tau_\mathrm{qp}^\mathrm{Nb} \simeq 150$\,ps and $\tau_\mathrm{qp}^\mathrm{Nb/Co} \simeq 50$\,ps. The temperature dependences $\tau_\mathrm{qp}(T)$ at $B = 10$\,mT are displayed in Fig. \ref{f4}(f). With a temperature  decrease from $0.975T_\mathrm{c}$ to $0.64T_\mathrm{c}$, $\tau_\mathrm{qp}(T)$ increases for all samples. For the as-grown Nb film this increase is by about an order of magnitude, while it is only a smaller factor for the Nb/Co hybrid structures.

Previously, it was argued that one can expect a rather strong temperature dependence of the quasiparticle relaxation time in cleaner samples \cite{Per05prb} in which $\tau_\mathrm{qp}(T)$ is expected to fit with a recombination model \cite{Doe97prb} predicting $\tau_\mathrm{qp}(T) \propto \exp[2\Delta(T)/k_\mathrm{B}T]$, where $\Delta(T) = \Delta(0)(1 - T/T_\mathrm{c})^{1/2}$ is the energy gap of the superconductor. Figure \ref{f4}(f) reveals that $\tau^\mathrm{Nb}_\mathrm{qp}(T)$ in the as-grown Nb film indeed fits to the recombination model \cite{Doe97prb} where the only adjustable parameter $\Delta(0)\approx 1.85 k_\mathrm{B}T_\mathrm{c}$ is slightly larger than
the BCS value $\Delta(0)\approx 1.76 k_\mathrm{B}T_\mathrm{c}$. We note that previously even larger deviations from the weak-coupling BCS limit were reported for Nb thin films \cite{Per05prb,Pro98prb}. In contradistinction, weaker dependences $\tau_\mathrm{qp}(T)$ are typically observed for Nb films of moderate quality and Nb-based hybrid structures \cite{Att12pcm}. Finally, in dirty superconductors one expects \cite{Kap76prb} a quasi-constant dependence of $\tau_\mathrm{qp}(T)$ \cite{Per05prb,Leo11prb}. In the order of magnitude, the deduced values of $\tau_\mathrm{qp}$ in our reference Nb and Nb/Co samples agree with the $\tau_\mathrm{qp}$ values known from literature \cite{Per05prb,Att12pcm}. At the same time, the second-power dependence of $\tau_\mathrm{qp}$ on the instability velocity stipulates a notable data scattering in the literature. In addition, $v^\ast$ depends not only on the sample quality and the dominating quasiparticle scattering mechanism but also on the cross-section of the constriction.

The data in Fig. \ref{f4}(c) and (f) for the Nb film decorated with Co stripes indicate that essentially higher vortex velocities can be achieved in the vortex guiding regime. Since the relaxation times in the Nb/Co film and in the pinning channels in the Nb film under Co stripes are not expected to differ substantially, the high values of $v^\ast_\mathrm{Nb/Co\,stripes}$ should imply a much narrower distribution of vortex velocities in the guiding regime. The assumption of different widths of the distribution of vortex velocities is consistent with the relation $v^\ast(0^\circ) < v^\ast(90^\circ)$ following from Figs. \ref{f2}(a) and \ref{f4}(b). Namely, one has to take into account that in the studied range of magnetic fields the vortex lattice is sparse and the depinning currents for vortex motion along and transverse the Co stripes differ by about a factor of four, see Fig. \ref{f4}(a). Accordingly, at $\alpha = 90^\circ$ most of the vortices are expected to move along the Co stripes where the order parameter is suppressed, while a minor fraction of interstitial vortices may still move in between the Co stripes. The weaker pinning between the Co stripes leads to higher velocities of interstitial vortices, that results in a broader distribution of vortex velocities \cite{Ada15prb}. In this physical picture, a large number of vortices contributes to the resistance, whereas only a small number of them triggers the instability. This is a peculiar feature of the local instability \cite{Bez19arx} whose fingerprint is a jump in a \emph{linear} section of the $I$-$V$ curve, exactly as we observe at $\alpha = 90^\circ$. This is distinct from the conventional LO framework for instability jumps occurrring in a \emph{nonlinear} section of the $I$-$V$ curve. The higher instability velocities in the as-grown Nb film and the Nb/Co bilayer than in the decorated Nb film with $\alpha = 90^\circ$ mean that it is the stronger pinning variation in the decorated sample that hinders a stable flux flow at high vortex velocities. Accordingly, the smallest value $v^\ast(0^\circ)\simeq300$\,m/s can be explained by the fact that the arrangement of vortices is least favorable for a coherent vortex dynamics, as the difference between the pinning strength at the Co stripes and between them is the largest in this case.

\section{Conclusion}
To summarize, we have studied the non-equilibrium superconducting state generated by high-velocity vortex motion in Nb films decorated with Co nanostripe arrays. A stabilization effect of vortex guiding on the flux transport has been observed. This effect is most pronounced when the Lorentz force acting on vortices is tilted at a small oblique angle with respect to pinning channels. The observation of a factor of five shorter quasiparticle relaxation times deduced in the regime of vortex guiding along the Co stripes as compared to Nb films covered with a continuous Co layer raises the question whether the relaxation times deduced from the current-voltage measurements on samples with uncorrelated pinning should be treated as upper-bound estimates. In all, our findings render the well-known vortex guiding effect to open novel prospects for investigations of ultrafast vortex dynamics.

\begin{acknowledgments}
The authors thank A.\,I. Bezuglyj and V.\,A. Shklovskij for useful discussions.
OD acknowledges the German Research Foundation (DFG) for support through Grant No 374052683 (DO1511/3-1). This work was supported by the European Cooperation in Science and Technology via COST Action CA16218 (NANOCOHYBRI).
\end{acknowledgments}


%

\end{document}